# A Modular Supersonic Ping Pong Gun

Mark French, Rajarshi Choudhuri, Jim Stratton, Craig Zehrung and Davin Huston

School of Engineering Technology

Purdue University

**Abstract**

A vacuum-powered device that shoots ping pong balls at high subsonic speeds has been used for physics demonstrations for more than a decade. It uses physics that are easily understood by students, even though its operation is not immediately intuitive. The addition of a pressure plenum and nozzle results in muzzle velocities exceeding Mach 1.5. Balls are readily fired through ping pong paddles and sheets of plywood up to 12.7mm (1/2 inch) thick. Popular reaction to the device indicates that it is an effective way to spark interest in physics and engineering.

**Introduction**

A subsonic ping pong gun (often called a ping pong Bazooka) has been used as a classroom demonstration tool as a means of demonstrating physical mechanisms that are dramatic and initially non-intuitve [1].The device is remarkably simple, consisting of a plastic tube with ends sealed using foil or tape and a small vacuum pump.

To prepare the gun, a ping pong ball is placed in the barrel and both ends are sealed using plastic tape or, less often, aluminum foil and then air is evacuated using the vacuum pump. To shoot, the seal at the breech end of the barrel is punctured, allowing air at atmospheric pressure to enter the barrel.  Since there is no air in the barrel to resist acceleration of the light ping pong ball, whose mass is about 2.5gm, initial acceleration is in the neighborhood of 5,000g.

As the ball moves down the barrel, the pressure differential must accelerate not only the ball, but a growing column of air in the barrel.  The increase in mass necessarily reduces acceleration as the distance from the breech to the ball increases [2].  The barrel is typically 2m,-3m long and it is worth noting that the mass of a 2m long column of air at standard temperature and pressure and a diameter of 40mm (the diameter of the ball), has a mass of about 3.1gm, more than that of the ball.

The last stage of the process is the ball leaving the barrel.  The ball should not fit tightly in the barrel, so a small volume of air moves past as it moves down the barrel.  When the ball reaches the muzzle, this bit of air preceding it is compressed between the ball and the muzzle seal.  The resulting pressure breaks the seal, leaving the muzzle open for the ball.  Figure 1 shows these steps in the firing event.

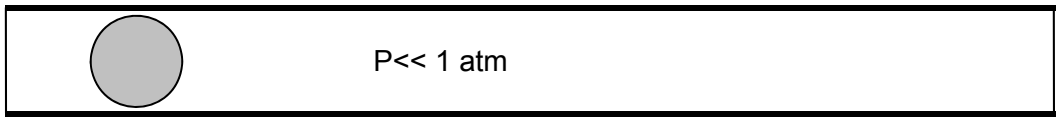

a) Tube is evacuated with end seals in place

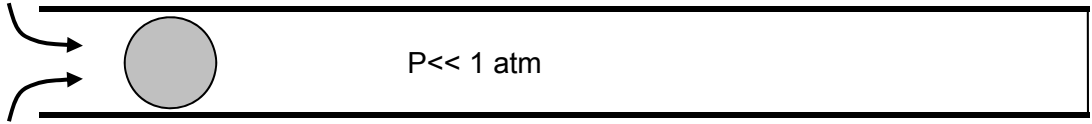

b) Seal is punctured and air rushes into vacuum

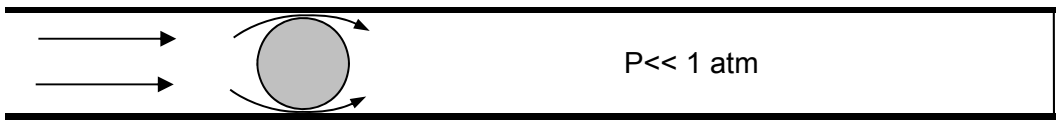

c) Pressure differential accelerates ball and some air leaks past

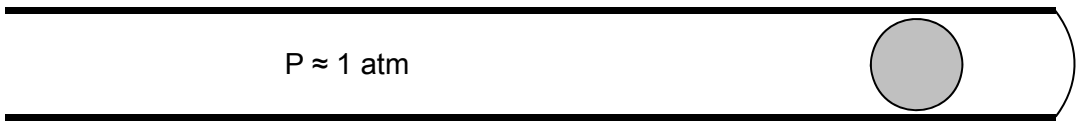

d) Air ahead of the ball is compressed and breaks muzzle seal

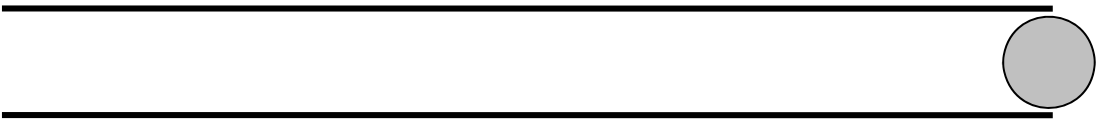

e) Ball exits through open muzzle

**Figure 1** – Operation of the Subsonic Ping Pong Gun

High speed video routinely shows the tape seal at the muzzle bulging out and bursting before the ball reaches it. The event is short enough that thousands of frames per second are needed to clearly capture it on video. Figure 2 shows a frame from high speed video showing the muzzle seal bursting. The light line superimposed over the images shows the location of the muzzle; the tape seal has clearly bulged, prior to bursting. The video was recorded at 8,000 frames/sec since it was difficult to observe the muzzle seal bursting at slower frame rates.

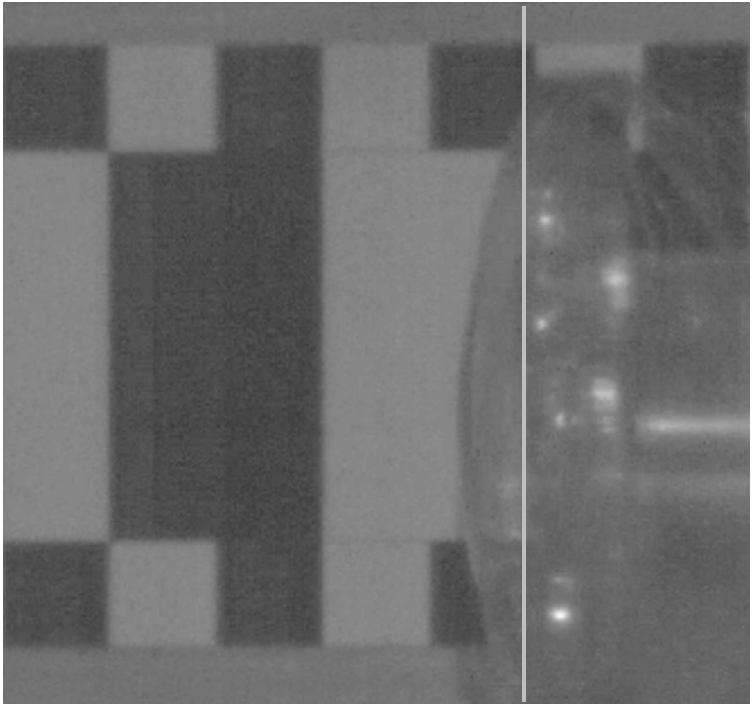

**Figure 2 –** Muzzle Seal Bursting

A simple analytical model can describe the basic operation of the subsonic gun, though the details of the internal flow are complex. An investigation of the subsonic gun correlated experimental results with an analytical model and used computational fluid dynamics (CFD) to model the first few milliseconds of the motion of the ball [3].

**The Supersonic Gun**

It seems self-evident that the challenge of increasing the muzzle velocity is a worthy one. The design of the device limits the velocity of the ball to that of the column of air behind it, so the goal is then to increase the velocity of air pushing the ball down the barrel. At the simplest level of analysis, increasing muzzle velocity requires increasing the energy available to the ball. One way to achieve this is to vent air into the barrel at higher than atmospheric pressure. This is analogous to the design of some supersonic wind tunnels sometimes called blowdown tunnels [4].

With this in mind, a pressure plenum was placed at the breech, with a burst disk separating the pressurized air volume from the vacuum in the barrel, as shown in Figure 3.

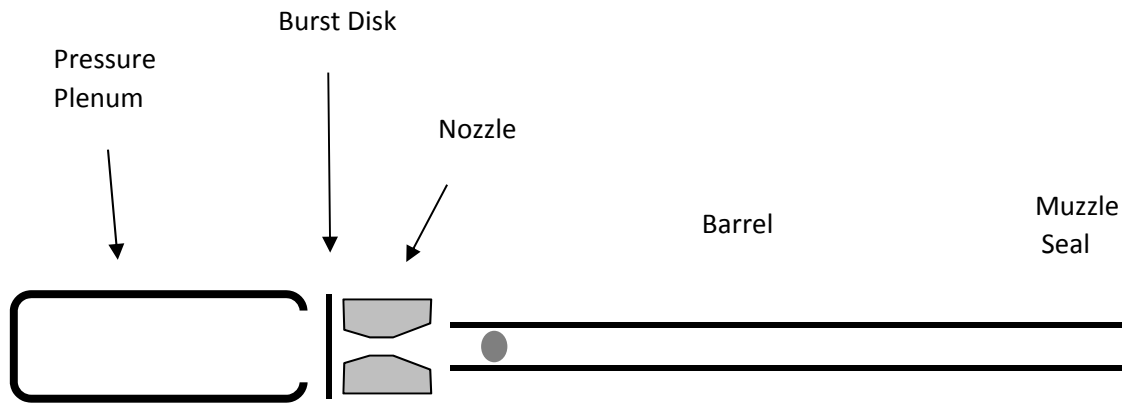

**Figure 3** – Supersonic Ping Pong Gun

The initial version of the device was made from readily available Schedule 80 PVC pipe and fittings. The burst disk was made from layers of clear plastic cold laminating film. After several experiments with remotely actuated devices to break the burst disk, we found the most reliable method was to simply increase supply pressure until the disk burst from overstress. Varying the number of layers of film gave some control over the burst pressure. The nozzle joining the pressure plenum to the barrel was machined from a solid rod of PVC with female threads to engage the treads on the end of the pressure plenum.  To ensure an air tight seal, the layered plastic sheet was simply wrapped around the end of the pressure tank and before it was threaded into the barrel.  Figure 4 shows the device ready to shoot.

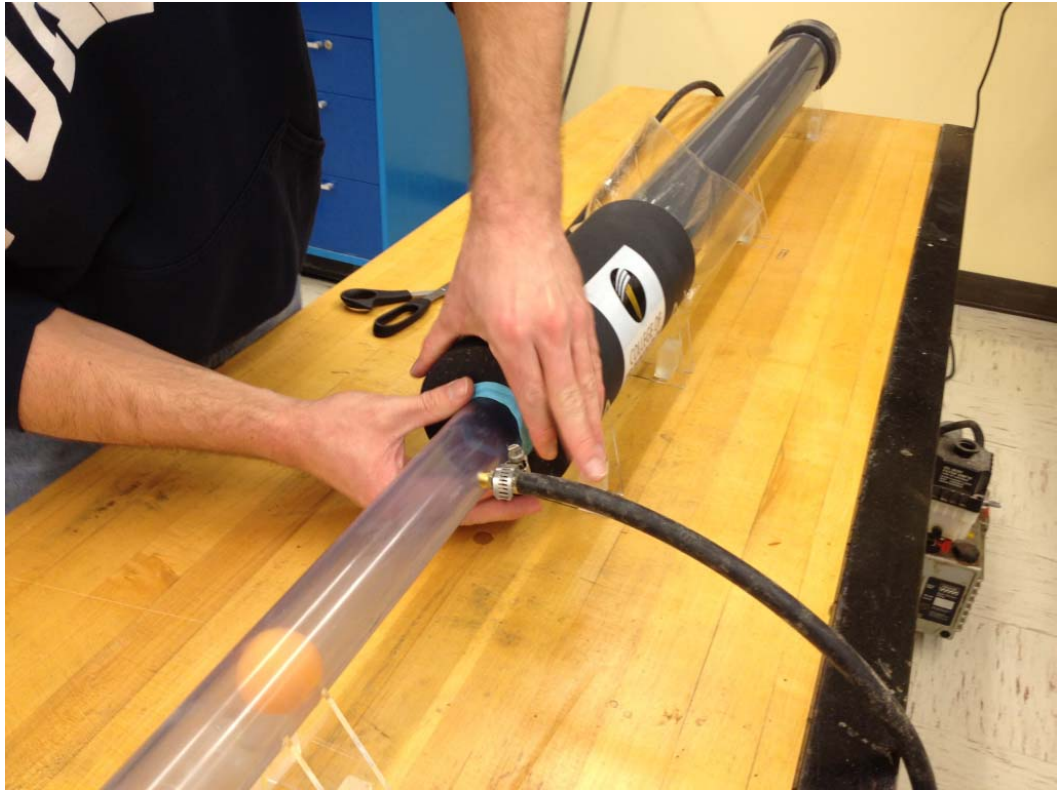

**Figure 4** – Initial Version of the Supersonic Ping Pong Gun

An important part of any supersonic wind tunnel is the nozzle between the high pressure air supply and the test section holding the model.  While looking very much like a venturi used to accelerate subsonic flow, its operation is fundamentally different for supersonic flow.  In this application, it is called a deLaval nozzle or a convergent-divergent nozzle (or, more simply, a C-D nozzle).  A 1-D analysis assuming isentropic flow shows that the Mach number downstream of the nozzle depends only on the geometry of the nozzle and the pressure ratio across it.

The initial nozzle design was kept simple, so that it could be manufactured easily on a lathe [5]. The pressure ratio approaches infinity as the barrel pressure approaches zero, but the nominal pressure in the barrel was typically about 1% of atmospheric. With a plenum pressure of 6-7 atmospheres, the pressure ratio was assumed to be in the range of 3000-6000.

The obvious problem with the supersonic wind tunnel analogy is that the barrel is almost completely blocked by the ball, the opposite of a wind tunnel where blockage due to the model is minimized. For reasons that will be described in detail below, modeling the operation of the device was not possible, so the effect of the nozzle needed to be evaluated experimentally.

In the second version of the device, the PVC pressure plenum and nozzle were replaced with machined aluminum components. The allowable pressure is much higher, and, since the operating pressure was not increased, the effective safety factor was increased. We found that the ping pong ball broke into pieces in the barrel when firing pressure exceeded about 825 kPa (120 psi), so there was no need to go to a higher pressure.

Additionally, the one piece nozzle was replace with an aluminum housing into which 3D printed nozzles could be inserted. This allowed us to test different nozzle geometries much more easily and the range of possible geometries was limited only by what shapes could be 3D printed. The pressure plenum and nozzle housing are now connected by bolted flanges that also hold a laser cut burst disk. This modular design allows changes to be made quickly while decreasing time to reset it for another shot and reducing the variation in muzzle velocity, shot to shot. This revised assembly is shown in Figure 5.

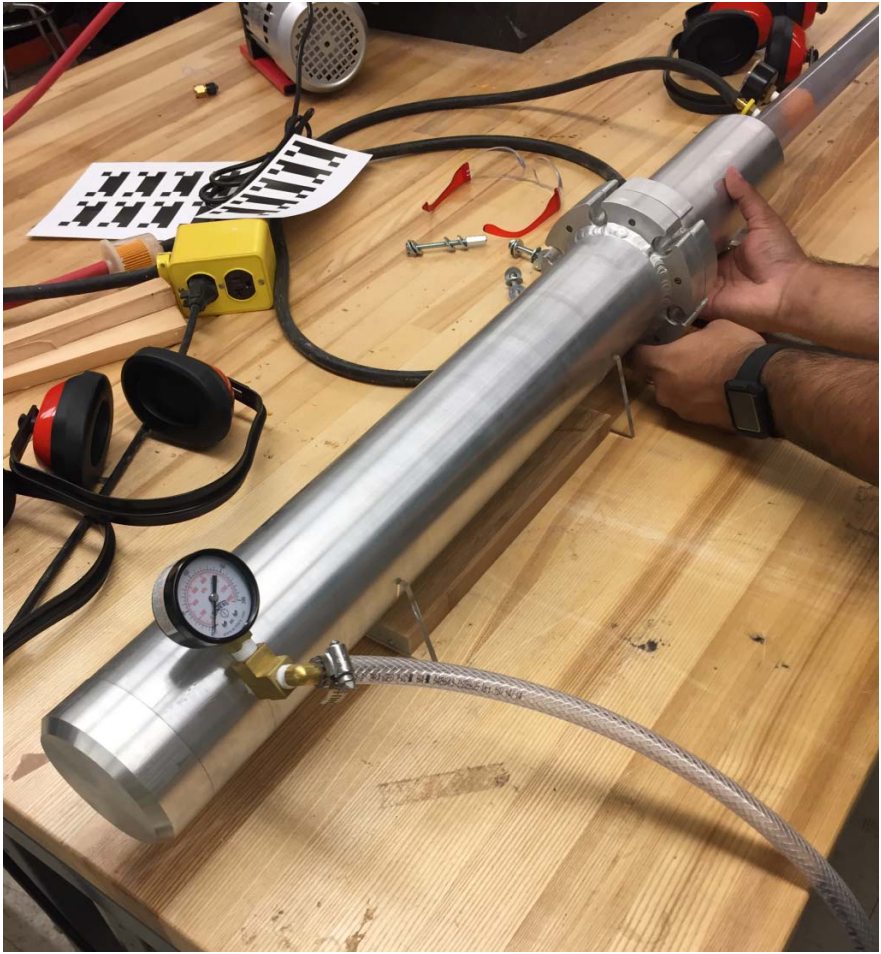

**Figure 5** – Pressure Plenum and Nozzle Housing

**Experimental Results**

We recorded the velocities from 30 tests with varying firing pressures and two different nozzle geometries.  One nozzle used a convergent-divergent cross section and the other was just a smooth transition from the larger diameter pressure plenum to that of the barrel.  The two cross-sections are shown in Figure 6.  The flow direction is from left to right.  The seat at the right edge of the nozzle is to accommodate the barrel.  The transition from the nozzle to barrel is smooth when the barrel is in place.

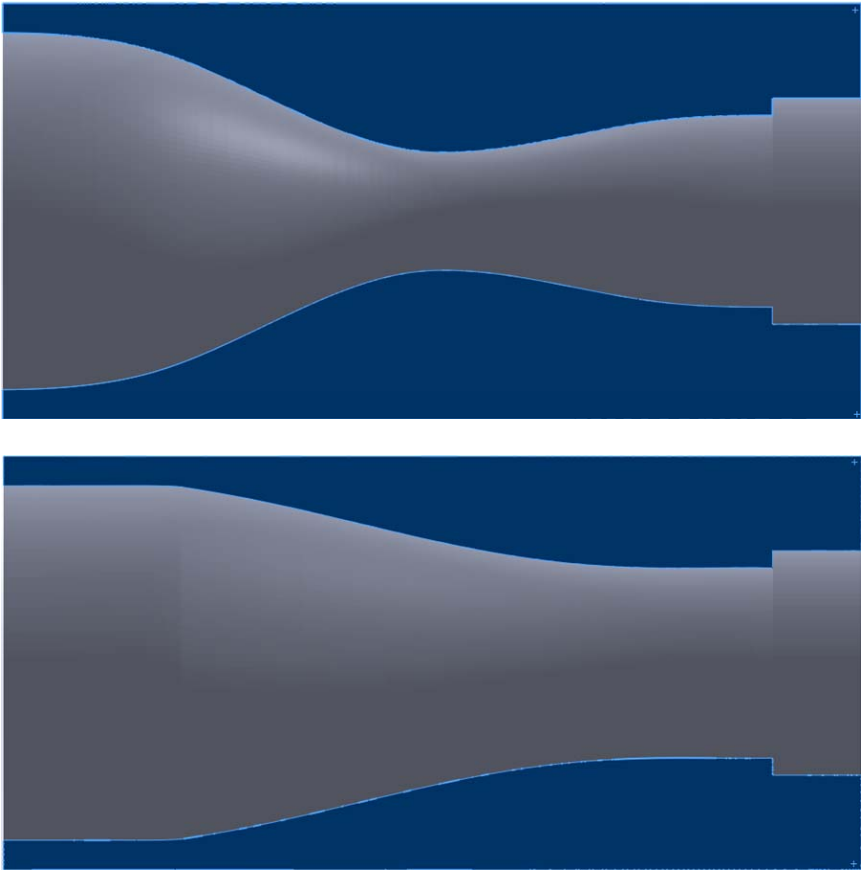

**Figure 6** – Two Nozzle Cross-sections

Making a reliable burst disk was one of the most challenging tasks in getting the supersonic gun to work. The final design was made from several layers of Xyron cold laminating film and cut to shape with a CNC laser. A light 'x' was cut into the top surface to create a stress concentration that would initiate failure at high pressure.

The hole created when the disk was overstressed by the pressure differential varies from shot to shot. Ideally, the ruptured burst disk moves completely out of the way to clear the path for the compressed air to enter the evacuated barrel. However, the rupture in the disk does not always reach the inner surface of the nozzle and the remaining plastic hinders flow.  Figure 7 shows two ruptured disks.  The disk on the left restricted flow much less than the one on the right.

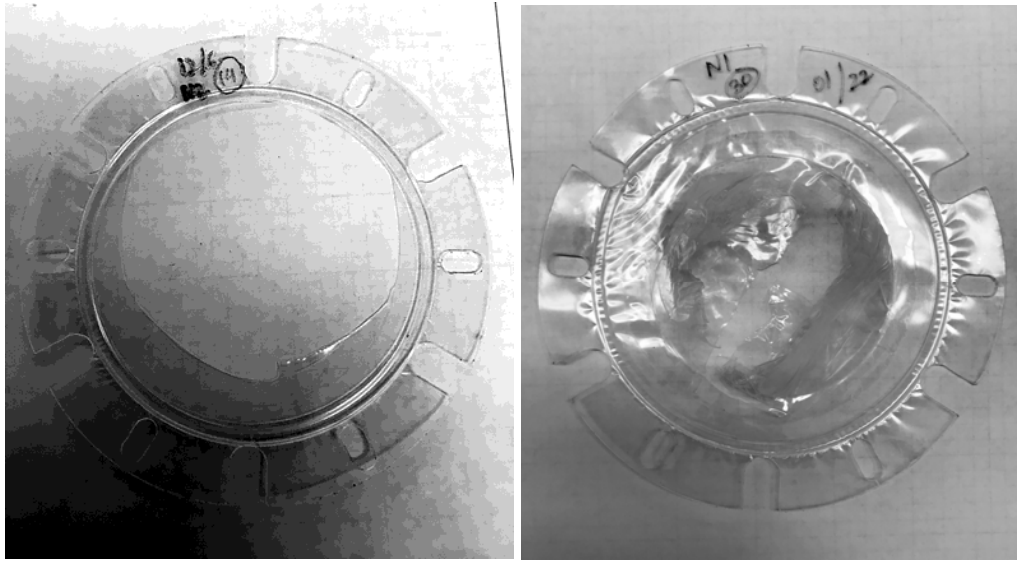

**Figure 7** – Ruptured Burst Disks

Variations in the failure pressure of the disks and in the passage opened by the failed disks gave varying muzzle velocities.  Figure 8 shows the muzzle velocities from a pool of 30 tests.  This includes two different nozzles and burst disks made from both three and four layers of film.

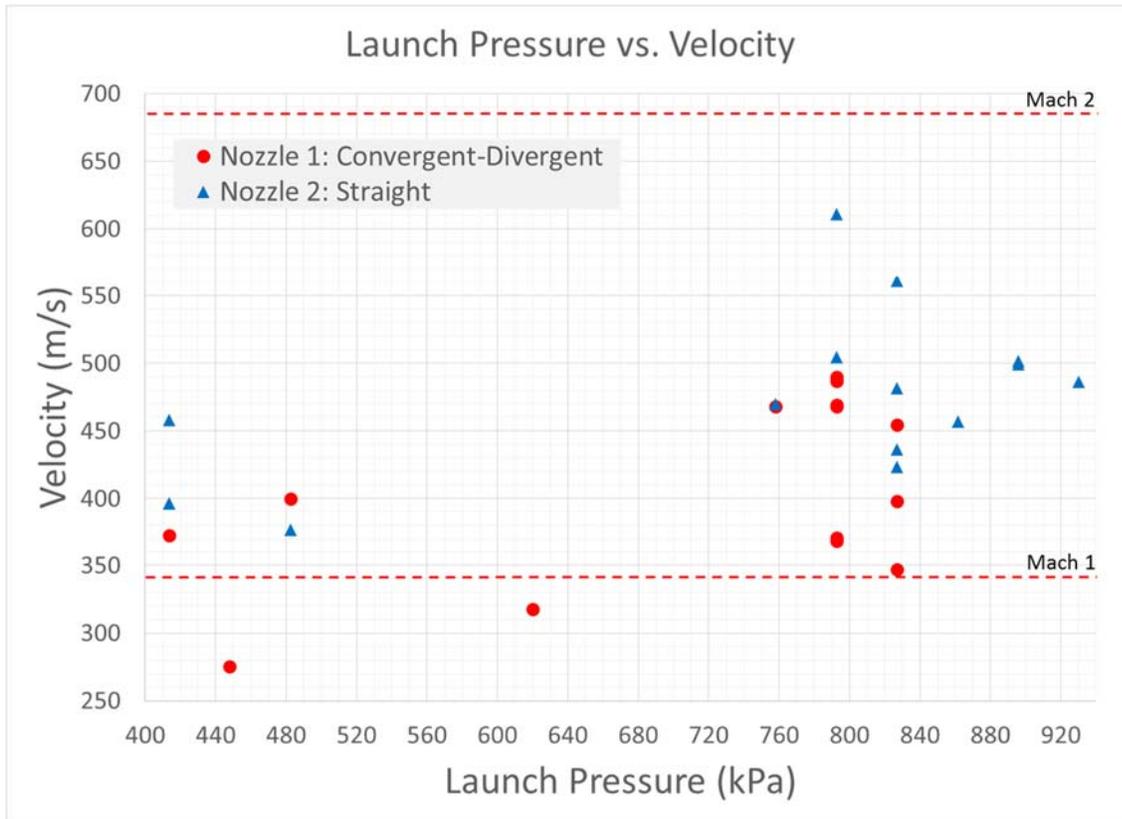

**Figure 8** – Muzzle Velocities

Generally, the straight nozzle gave higher muzzle velocities than did the convergent-divergent nozzle. This suggests that the analogy of the supersonic ping pong gun to a supersonic wind tunnel is not a perfect one. In order to full characterize the operation of the device, an analysis capable of modelling the transient compressible flow and its effect on the ball is probably needed.

**Firing and Impact**

Of course, the higher velocity of the supersonic ping pong gun gives the ball much higher kinetic energy than the subsonic gun. Since the ball is moving in a column of air, the kinetic energy of the air is a likely contributor to the energy dissipated at impact. We experimented with different targets, but have settled on a few useful ones, all simple and inexpensive. When the goal is to show the energy of the ball to students, the gun is often used to shoot through empty aluminum drink cans. It is routine to shoot through five cans as shown in Figure 9. For safety, there needs to be a way to capture the cans and the pieces of the ball – which breaks into small pieces after impact. A few sheets of corrugated cardboard or a large box with Styrofoam inserts works well. The first can is typically placed about 100mm from the muzzle. For comparison, the subsonic gun can only shoot through two cans.

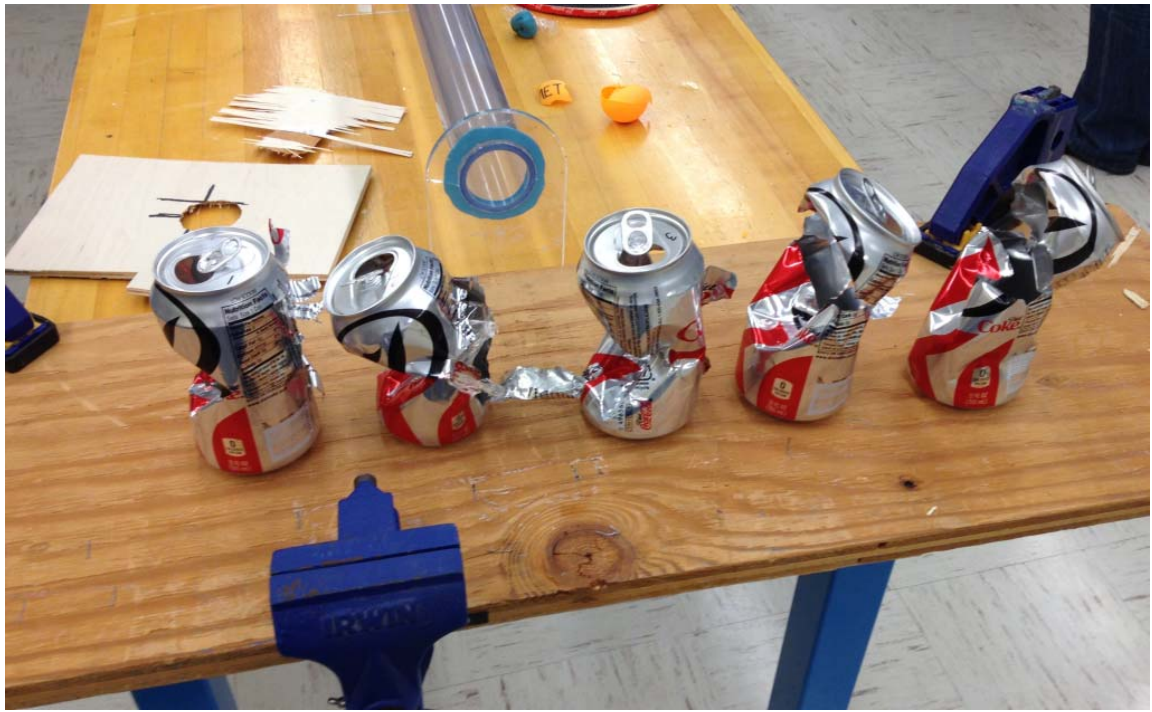

**Figure 9** – Empty Drink Cans after a Typical Test

The other useful targets are sheets of plywood or oriented strand board (OSB), of the type typically used for subflooring in houses. At the highest muzzle velocity, the ball can typically penetrate 12.7mm (1/2 inch) sheets of plywood and 15.88mm (5/8 inch) thick OSB. If the ball leaves the barrel intact, the resulting hole can be reasonably clean. Figure 10 shows the effect of four successive shots on a piece of OSB.

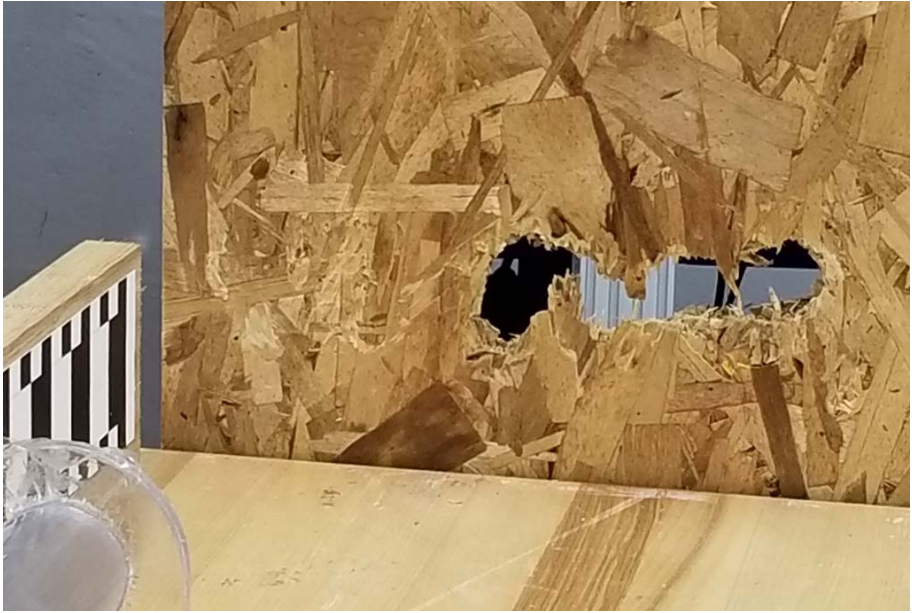

**Figure 10** – 15.88mm (5/8 inch) OSB Target

The mechanical impedance of the target greatly affects the impact event.  In order to penetrate plywood OSB, the target piece has to be clamped securely so that impact forces are maximized.

When the target was a piece of steel sheet, results differed in two ways.  The first was that the ball didn't shatter, as it did with other targets.  Rather, the hemisphere that impacted the target buckled, suggesting that the high strain rate softened the otherwise brittle plastic.  The second was that the resulting dent in the sheet, while circular, was flat-bottomed rather than a convex cavity as one might have expected.  Figure 11 shows the result of a test shot.  The front hemisphere of the ball has plastically deformed and, while hard to see in a picture, has a slightly darker color than it had before the test.  The circular, flat-bottom dent is also clearly visible.

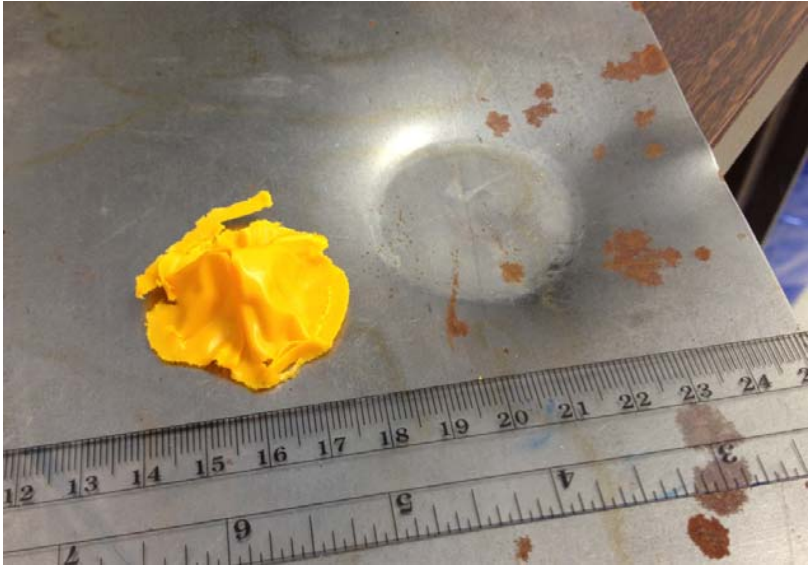

**Figure 11 –** Impact with Steel Sheet

Ping pong balls are light and relatively fragile. It is not surprising, then, that it is possible to break the ball in barrel if the acceleration is too high. In practice, plenum pressures above about 825 kPa (120 psi) generally cause structural failure during firing. The gun still fires, but the ball leaves the gun as a collection of fragments. These fragments still have kinetic energy and can sometimes still penetrate a target board.

To investigate the effects of increased mass, a series tests were conducted with a water-filled ball. Water was injected through a small hole drilled in the ball and the hole was then sealed with super glue. The resulting impact transferred much more energy to the target. For example, the steel sheet shown in Figure 11 was torn from its clamps and heavily deformed.

**Analytical Challenges**

In spite of its simplicity, the supersonic ping pong gun poses major analytical challenges. It violates some of the simplifying assumptions used to make computational fluid dynamic (CFD) modelling more tractable [6]. Among these are:

- Highly time dependent flow – Firing time is measured in milliseconds so time steps must be very short.
- Moving boundary conditions – The position of the ball changes quickly and displacement is large compared to the diameter of the barrel, so the grid cannot just distort in order to accommodate motion. Rather, it must add grid points continuously as the ball moves. As velocity increases, the rate of grid growth must increase with it.
- Motion due to calculated pressures – The motion of the ball is not known beforehand, so its position must be recalculated at each time step based on pressures calculated across the surface of the ball.
- High pressure gradients – At the instant of firing, the pressure on one side of the ball is close to zero, while that on the other side might exceed five atmospheres.

In principle, the most general numerical implementation of the Navier-Stokes equations should be able to address all these challenges. However, modeling all these phenomena within commercially available CFD packages is likely to be difficult.

**Second Order Physical Observations**

In addition to the primary purpose of accelerating a ping pong ball to supersonic speed, we observed some interesting secondary phenomena. The first of these was oscillation of the ball in the barrel as the vacuum pump evacuated the barrel. As pressure was reduced, the ball oscillated back and forth around the vacuum port until the barrel was nearly at a vacuum. This only occurred with the larger of the two pumps we used, suggesting that flow rate past the ball was a factor. In practice, the ball stopping indicated the barrel was pumped down far enough for the device to shoot [7].

A second, and perhaps more intuitive affect was that the gun recoils when it fires. At first glance, it would appear that air rushing in to fill a vacuum might not generate a net force. However summing forces along the barrel axis clearly shows a net backward force. High speed video showed that observable recoil began before the ball left the barrel. As an extension, we then found that securing the gun to prevent recoil also increased the muzzle velocity.

**Popular Reception**

Normally, popular response to an effort like this would not be part of a technical article. However, a primary use of the ping pong gun is as a teaching tool, so an enthusiastic response is an important measure of its success. The initial version of the device was described in a short post on ArXiv and the popular response was immediate. Within a few days, the authors were contacted by news and media from around the country and, eventually, including Great Britain. Articles and video reports began to appear and reporters came to the university to record interviews. Later, the BBC sent a team, including a well-known presenter, to our campus to record an episode of a series documentary devoted to extreme weather events. Two of us (Stratton and Zehrung) found a way to make artificial hail stone which they then fired at targets using the ping pong gun. The supersonic ping pong gun became the subject of a segment on Mythbusters and interest peaked when two of the devices were demonstrated on a science segment of Late Night with Jimmy Fallon. One of us (French) has a YouTube channel devoted to short informational videos for engineering and engineering technology students [8]. The video describing the supersonic ping pong gun quickly became the most popular video on the channel and has accumulated more than 754,000 views.

That this device has attracted wide interest strongly suggests its utility as teaching tool. The basic principles underlying its operation are not hard to understand, but it is difficult to explain its operation in detail. This, combined with the dramatic effects of a ping pong ball hitting a target can create a unique opportunity to teach physics and engineering.

**Future Directions**

After the initial reception of the supersonic ping pong ball gun, it is not surprising that that others are conducting their own experiments. At least one of these is being tested with helium rather than air as the working fluid [9]. Several options for future work suggest themselves:

- Changes in working fluid from very light (Helium) to very heavy (Sulphur Hexafluoride)
- Modified projectiles – ping pong ball with heavier wall to survive launch acceleration
- Refined nozzle geometries – reduction of internal losses could and tailoring it more carefully for transient compressible flow could make the device more efficient
- Development of a more consistent burst disk design

**Conclusions**

A simple addition to the subsonic ping pong ball gun reliably gives a supersonic muzzle velocity.  A modular implementation of the device allows easy changes of the burst disk and of the nozzle.  Tests using two different nozzle geometries suggest that a throat, like that used on supersonic wind tunnels, reduces the muzzle velocity.  More testing and an accurate numerical model may be needed in order to generalize this result.

The supersonic ping pong gun generated a surprising popular response, attracting attention from magazines, web sites and television shows.  This suggests that the device can be very effective in interesting students in physics and engineering.

**Acknowledgements**

We wish to thank John Huhn of Motion Engineering for his generous assistance in recording impacts as very high frame rates.  The images he helped us record have reached around the world.  We also wish to thank Sean Cline for his assistance in fabricating parts for the pressure plenum and the nozzle housing.

1. J. Cockman, "Improved Vacuum Bazooka," *The Physics Teacher,* vol. 41, pp. 246-247, 2003.
2. E. Ayars and L. Buchholz, "Analysis of the Vacuum Canon," *The American Journal of Physics,* vol. 72, no. 7, pp. 961-963, 2004.
3. R. M. French, V. C. Gorrepati, E. Alcorta and M. J. Jackson, "The Mechanics f a Ping Pong Gun," *Experimental Techniques,* no. January/February, pp. 25-30, 2008.
4. A. Pope and K. L. Goin, High-Speed Wind Tunnel Testing, Krieger, 1978.

5. M. French, J. Stratton and C. Zehrung, "A Supersonic Pring Pong Ball Gun," in *Proceedings, Vibration Institute Annual Training Conference*, Indianapolis, 2017.

6. R. H. Pletcher, J. C. Tannehill and D. A. Anderson, Computational Fluid Dynamics and Heat Transfer, 3rd Edition, CRC Press, 2011.

7. R. French, C. Zehrung and J. Strattion, "A Supersonic Ping Pong Gun," ArXiv, 2013.

8. R. French, "Youtube," [Online]. Available: youtube.com/user/PurdueMET. [Accessed 4 February 2018].

9. Davekni, "YouTube," [Online]. Available: https://youtu.be/Z52yCL3tSGQ. [Accessed 4 February 2018].